# Avalanche photodiodes based on MoS$_2$/Si heterojunctions

Oriol Lopez-Sanchez[1], Dumitru Dumcenco[1], Edoardo Charbon[1,2*], Andras Kis[1*]

[1]*Electrical Engineering Institute, Ecole Polytechnique Federale de Lausanne (EPFL), CH-1015 Lausanne, Switzerland*

[2]*Department of Microelectronics, Delft University of Technology, Mekelweg 4, 2628 CD Delft, The Netherlands*

*Correspondence should be addressed to: Andras Kis, andras.kis@epfl.ch and Edoardo Charbon, edoardo.charbon@epfl.ch

**Avalanche photodiodes (APDs) are the semiconducting analogue of photomultiplier tubes offering very high internal current gain and fast response. APDs are interesting for a wide range of applications in communications[1], laser ranging[2], biological imaging[3], and medical imaging[4] where they offer speed and sensitivity superior to those of classical p-n junction-based photodetectors. The APD principle of operation is based on photocurrent multiplication through impact ionization in reverse-biased p-n junctions. APDs can either operate in proportional mode, where the bias voltage is below breakdown, or in Geiger mode, where the bias voltage is above breakdown. In proportional mode, the multiplication gain is finite, thus allowing for photon energy discrimination, while in Geiger mode of operation the multiplication gain is virtually infinite and a self-sustaining avalanche may be triggered, thus allowing detection of single photons[5]. Here, we demonstrate APDs based on vertically stacked monolayer MoS$_2$ and p-Si, forming an abrupt p-n heterojunction. With this device, we demonstrate carrier multiplication exceeding 1000. Even though such multiplication factors in APDs are commonly accompanied by high noise, our devices show extremely low noise levels comparable with those in regular photodiodes. These heterostructures allow the realization of simple and inexpensive high-performance and low-noise photon counters based on transition metal dichalcogenides.**

Ultrasensitive photon detection and single-photon detection capabilities are at the heart of numerous applications that include cryptography[6], laser ranging and LiDAR[2], three-dimensional imaging[7,8], biological imaging[3] and single-molecule detection[9]. These applications exploit the unique performance of avalanche photodiodes (APDs), optoelectronic devices with high internal gain, capable of detecting single photons. APDs consist of p-n junctions operated in the reverse bias regime. Incoming photons are converted into charge carrier pairs which are accelerated in the high electric field of the p-n junction, creating a cascade of carrier pairs through the process of impact ionization. APDs can be operated in proportional or Geiger modes. In proportional mode, APDs are biased below the breakdown voltage and their output current is related to the incident optical power. By biasing the device well above the breakdown voltage, incoming individual photons can trigger an avalanche breakdown, resulting in current spikes. The light intensity can then be obtained by counting the number of output pulses within a given time period. Today, proportional mode APDs or Geiger mode APDs, also known as single-photon avalanche diodes (SPADs) can be fabricated in low-cost complementary metal-oxide semiconductor (CMOS) technology, thus allowing their use in consumer electronics, however their adoption is still limited to specific applications due to reduced spectral range and small active surfaces.

Because two-dimensional dichalcogenide semiconductors could in principle be produced on large scales using liquid scale exfoliation[10] or CVD[11–13], their use as one of the components of APD devices could result in a dramatic reduction of price and would bring APDs with their unique capabilities into mainstream consumer electronic devices. A typical example of layered transition metal dichalcogenide (TMD) semiconductors[14,15] is molybdenum disulphide ($MoS_2$). Bulk TMD crystals are composed of stacks of weakly interacting single-layers that can be extracted using for example the adhesive tape-based exfoliation technique[16] developed for the preparation of graphene[17], The presence of a band gap of ~1.8 eV in $MoS_2$[18] has allowed its use as the basic building block of room-temperature field-effect transistors with high on/off current ratios[19], logic circuits[20], amplifiers[21], ring oscillators[22] and gigahertz-range transistors[23]. Because of quantum confinement in the two-dimensional limit, a transition from an indirect band gap to a direct band gap occurs in $MoS_2$ and other TMD semiconductors[18,24–26]. This makes monolayer $MoS_2$ and other 2D TMD materials interesting for applications in optoelectronic devices such as phototransistors[27], ultrasensitive photodetectors[28] or LEDs[29]. The functionality of these devices can be further enhanced by combining them with other 2D[30,31] or 3D[32] materials into vertical heterostructures in order to realize p-n junctions. The 2D/3D heterostructure could be especially interesting for the realization of avalanche photodiodes. Here, the 2D semiconductor such as $MoS_2$ could act as a light absorbing material while the region of the 3D semiconductor close to the abrupt junction would act as a charge multiplication region.

We fabricated our avalanche photodiodes by transferring $MoS_2$ grown on atomically smooth sapphire[33] using the CVD method[12,13] onto a pre-patterned substrate in the form of p-type doped Si[32,34] covered with $SiO_2$, as shown in Fig. 1. Prior to transfer, windows are opened in $SiO_2$ and the underlying Si is passivated with hydrogen[35]. The resulting abrupt p-n heterojunction diodes are capable of light emission and can also operate as solar cells[32]. The current as a function of bias voltage for a representative device is shown on Fig. 2a. The dark current ($I_{dark}$), follows a bias voltage dependence typical of p-n diodes, with the current under reverse bias voltages mostly due to drift, diffusion, thermal generation-recombination and tunneling[36]. Exposing the same device to light ($\lambda$ = 633nm, incoming illumination power $P_{inc}$ = 54.6 nW) results in a voltage-dependent photocurrent ($I_{ph}$) which increases with increasing reverse bias voltage until breakdown (not shown here).

From this data, we can calculate the multiplication factor M for the avalanche photodiode using the expression[37] $M = (I_{ph} - I_{dark})/I_{M=1}$ where $I_{M=1}$ is the unity gain photocurrent. We determine $I_{M=1} = 2.3 \times 10^{-10}$ A by measuring the photocurrent for zero bias, as we do not expect to see charge carrier multiplication in this regime of operation. The plot of the multiplication factor, shown on Fig. 2a. shows that our device can reach $M$ values higher than 1000. Our device can already show a multiplication value $M$ = 50 for a bias voltage as low as −1 V.

The proposed band structure of the device is shown on Fig. 2b and is typical of type-II abrupt heterojunctions[38]. Characteristic of the band structure are the band offsets for the conduction band of $\Delta E_C$ = 200 meV and valence band ($\Delta E_V$ = 900 meV), due to a difference in electron affinities[39] and band gaps of Si and $MoS_2$. The application of a reverse bias can further increase the discontinuities in the conduction and valence bands and is expected to result in a strong internal electric field.

We also characterize the responsivity of our APD. On Fig. 3. we show the photocurrent $I_{ph}$ and responsivity $R$ defined as $R = I_{ph}/P_{inc}$ recorded for different illumination powers, acquired while the device is operated at a bias voltage $V$ = − 40 V. At low illumination powers ($P_{inc}$ = 12.6 nW), the device shows a responsivity $R$ = 2.2 A/W. Increasing the light intensity results in a sublinear increase of the photocurrent, with the responsivity reaching 50 mA/W for $P_{inc} = 13.3$ µW

While APDs in general can offer high multiplication factors, their use as optical detectors can be limited by their noise performance. The randomness of the impact ionization process results in fluctuations of the multiplication factor, expressed as the excess noise factor $F$. In the absence of additional noise sources, the signal to noise ratio in APDs is $F$ times worse than in standard photodetectors ($M = 1$) with the same quantum efficiency[40]. The excess noise factor $F$ is related to the standard deviation of the multiplication $\sigma_M$ as[40]:

$$F = 1 + \frac{\sigma_M^2}{M^2} \quad (1)$$

Excess noise in APDs is generally explained in terms of McIntyre formula[37,41] giving the relationship between noise in APDs, the multiplication and the ionization ratio $k = \alpha_h/\alpha_e$, where $\alpha_h$ and $\alpha_e$ are ionization coefficients for electrons and holes, respectively. Extreme values of $k$ ($k \ll 1$ for devices based on electron multiplication or $k \gg 1$ for devices based on hole multiplication, implying that only one type of charge carriers undergoes multiplication), are desired because they offer low noise performance[40] with minimal $F$ values of 2 achieved in Si APDs[42].

In order to determine the excess noise factor of our APD, we first measure the APD current under constant illumination intensity as a function of time in order to determine the standard deviation of multiplication $\sigma_M$. Results for different bias voltages corresponding to different multiplication factors are shown on Fig. 4a. The device output is relatively stable up to bias voltages ~ −25 V, with long-term fluctuations starting to appear at higher bias voltages. The extracted excess noise factor, plotted on Fig. 4b however shows exceptionally low values with a maximum $F = 1.147$ at $M = 960$ and small variation across the measured range of multiplication values. Such a small variation is indicative of extreme ionization ratios and is lower than the minimum excess noise value achieved in Si APDs[42]. We note however that the McIntyre formula for noise in APD's and its result are based on the assumption of a uniform electric field in the multiplication region. The proposed band diagram of our device however implies the presence of a strongly varying electric field in the heterostructure and a new theoretical model will have to be developed in the future providing a link between excess noise and multiplication. Based on the strong asymmetry of our device, with the n-type region having an atomic scale thickness and the bulk p-Si region, we can expect that holes are much more likely to be multiplied in our device than electrons. The extremely low values of excess noise show that hybrid heterostructures based on the combination of 2D and 3D semiconductors with their highly inhomogeneous built-in electric fields offer a new way of realizing low-noise and low-cost avalanche photodetectors.

To summarize, we have fabricated avalanche photodiodes based on a hybrid heterostructure composed of CVD-grown n-MoS$_2$ and p-Si. Out devices show multiplication of photogenerated carriers, with a multiplication factor of 50 for a bias voltage as low as -1 V and exceeding 1000 for -45 V. The salient feature of our device is a very low excess noise, showing a very small degradation of the signal-to-noise performance at high gains. Together with the simplicity of the device design and increasing availability of large-area grown MoS$_2$, our device could open the way to the fabrication of inexpensive and low-noise APDs.

## METHODS

All measurements are performed in room temperature and under vacuum. Single layers of MoS$_2$ are grown using the reaction between MoO$_3$ and sulphur in a flow of ultrapure argon gas at atmospheric pressure and a temperature of 700 °C, using c-plane sapphire as the growth substrate[33]. MoS$_2$ monolayers were transferred[43] onto p-type silicon substrates with a resistivity of 0.1-0.5 Ωcm, corresponding to a boron doping level between $3\times10^{16}$ and $3\times10^{17}$ cm$^{-3}$, covered by 100 nm thick layer of thermal SiO$_2$ with patterned holes from 2 μm x 2 μm up to 5 μm × 5 μm. Windows in in SiO$_2$ are opened using 7:1 buffered oxide etch, resulting in sloped sidewalls. The initial etching step was followed 1 minute 1% HF etch in order to remove the native oxide and passivate the Si surface[35]. For electrical characterization, we use a gold electrode deposited on MoS$_2$ and large-area electrodes in direct contact with the p-Si substrate. A 30 nm thick HfO$_2$ layer is deposited on top of the device in order to encapsulate it. The device is illuminated using a 633 nm HeNe laser. Electrical measurements are carried out using a National Instruments DAQ card and Stanford research SR560 current preamplifier.

## ACKNOWLEDGEMENTS

Device fabrication was carried out in part in the EPFL Center for Micro/Nanotechnology (CMI). Thanks go to Zdenek Benes for technical support electron-beam lithography. We thank Jacopo Brivio and Simone Bertolazzi for technical help with the MoS$_2$ transfer and Yen-Cheng Kung for help with mask design and fabrication. This work was financially supported by a Swiss SNF Grant 153298 and Swiss SNF Sinergia Grant no. 147607.

**Author contributions**

OLS performed device fabrication and characterization. DD worked on CVD growth of MoS$_2$. EC initiated the research. AK supervised and co-ordinated the work. All authors performed work on manuscript preparation.

**FIGURES**

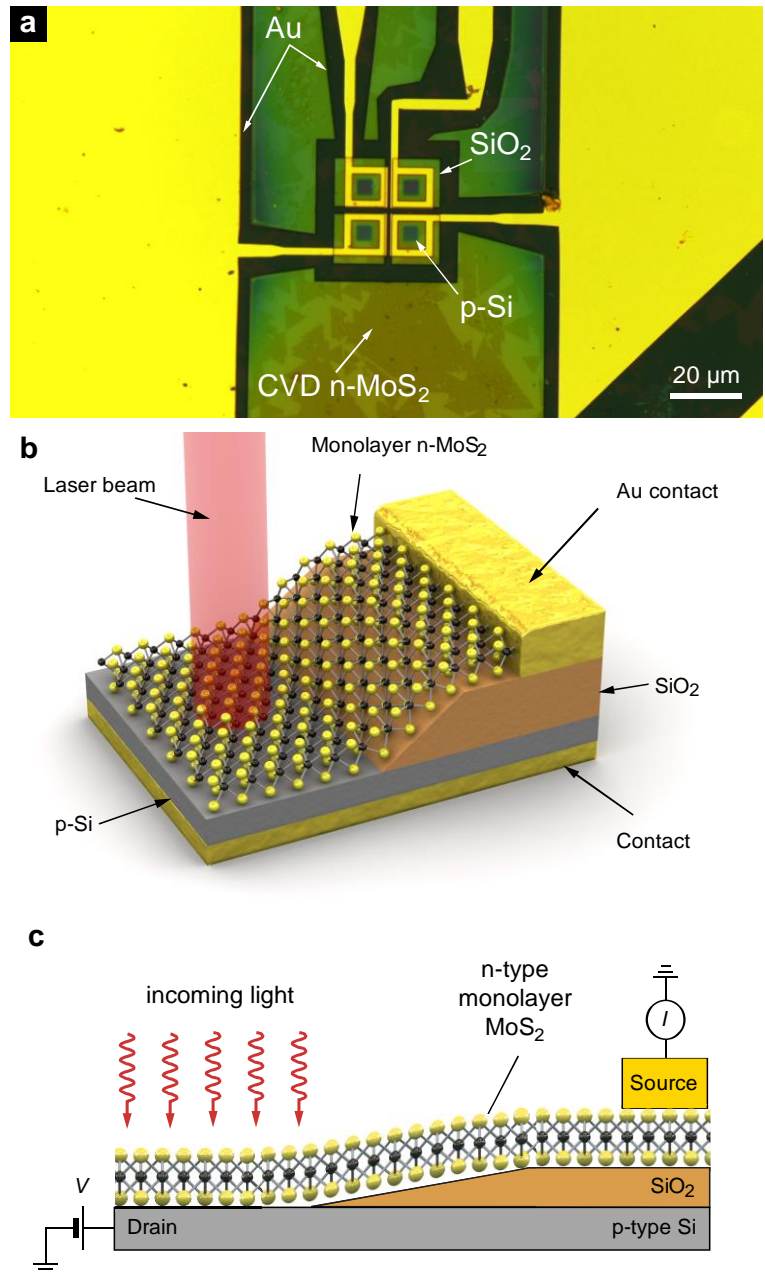

**Figure 1. Avalanche photodiode based on a MoS$_2$/Si vertical heterostructure. a,** Optical image of the device. **b,** Three-dimensional schematic view of the device. CVD-grown n-MoS$_2$ is transferred across a window opened in SiO$_2$ exposing underlying p-Si **c,** Cross-sectional view of the device structure together with the electrical connections used to perform the electrical characterization.

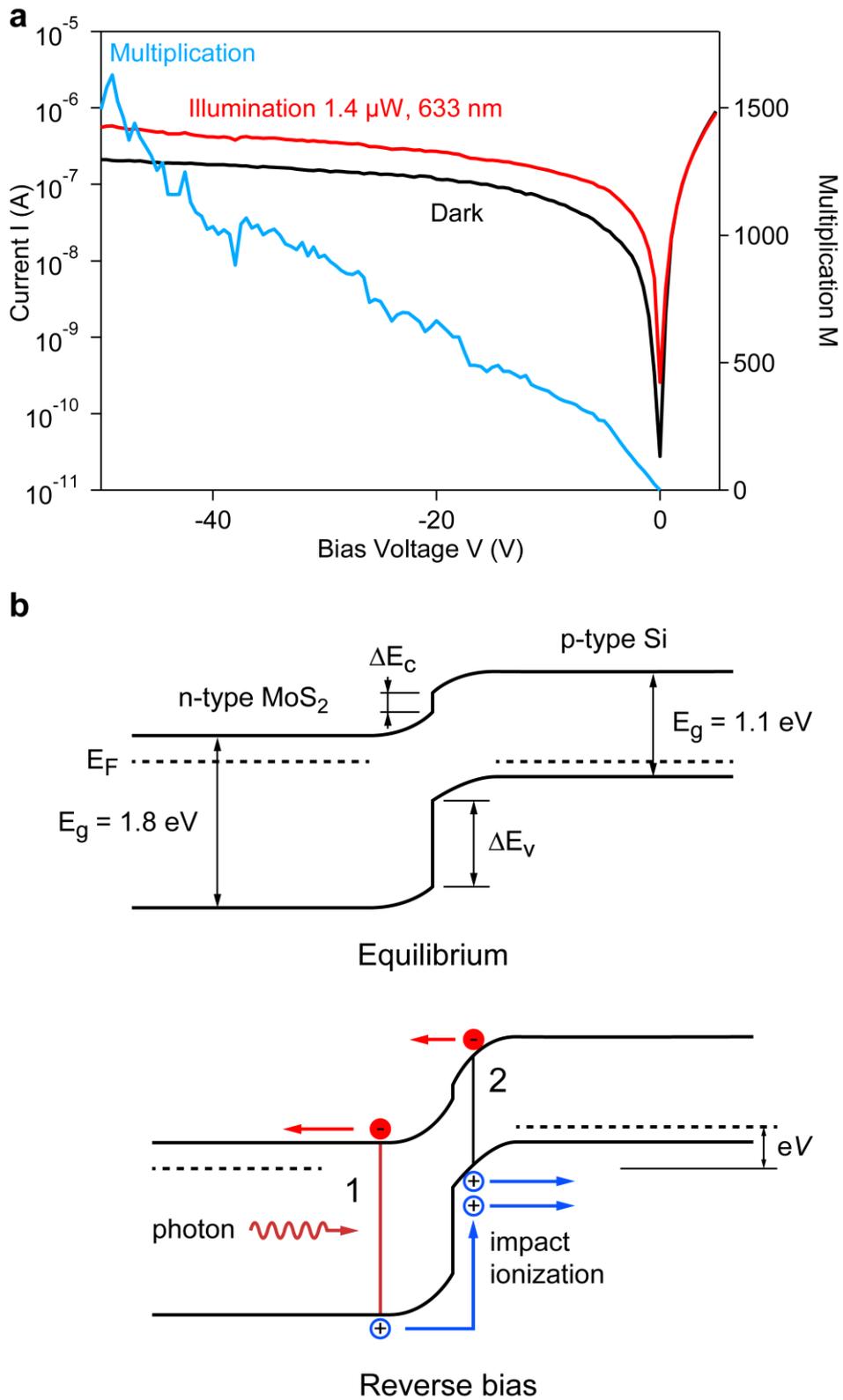

**Figure 2. Characterization of the MoS₂/Si APD. a,** Current voltage (I-V) characteristics of the APD (black line: dark, red line: under illumination with a 633 nm laser and incoming light power of 1.4 µW. **b,** Band diagram of the MoS₂/Si heterostructure in equilibrium. The absorption of a photon creates an electron hole pair. Under reverse bias, the resulting electrons and holes are accelerated under the influence of the electric field. Impact ionization generates a second electron-hole pair giving rise to multiplication.

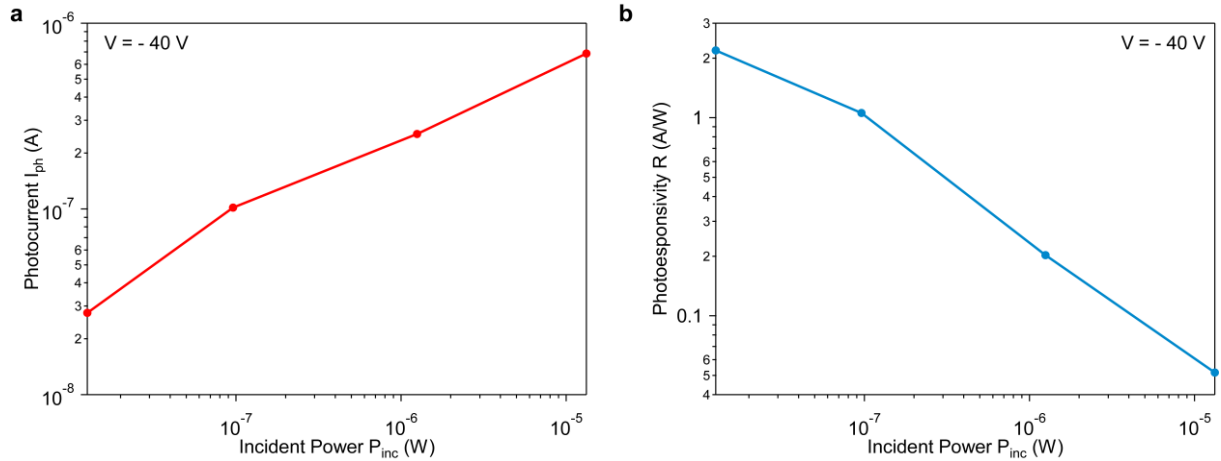

**Figure 3. APD response. a,** Photocurrent measured as a function of illumination power using a λ = 633 nm laser source for a bias voltage V = −40 V. **b,** Photoresponsivity of the MoS$_2$ phototransistor, showing a photoresponsivity of 2.2 A/W for an illumination power of 12.6 nW and shows a decrease with increasing illumination intensity.

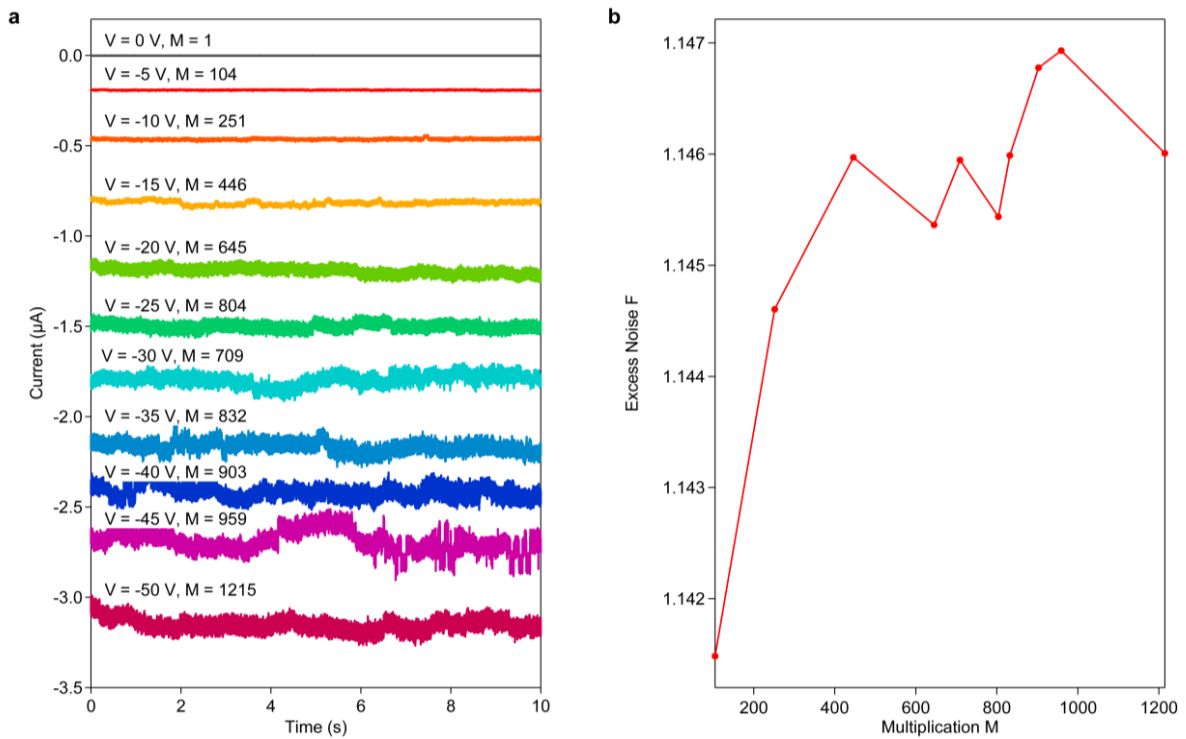

**Figure 4. APD noise characteristics. a,** APD current under constant illumination power of 101 nW using a λ = 633 nm laser source recorded as a function of time for different bias voltages and multiplication factors. At moderate bias voltages, the device output is relative stable with long-term fluctuations starting to appear at reverse bias voltages higher than 25 V. **b,** Excess noise factor *F* extracted from traces in (a) showing a very small variation with multiplication and very little degradation of SNR with respect to photodiodes.

8